# Force, quantum mechanics and approximate energy eigenstates

C. Das, Department of Chemistry, B. C. College, Asansol 713 304, India
and
*K. Bhattacharyya, Department of Chemistry, University of Calcutta, Kolkata 700 009, India

Abstract
The prevalent role of force in traditional quantum mechanics is outlined, with special reference to approximate calculations for stationary states. It will be explored how far this force concept can be made useful in the concerned area. The basic idea is to differentiate the Schrödinger stationary equation once. Thus, one can eliminate the unknown energy as well, and then examine how a force-based approach can be beneficial in providing quickly the nodal information and in assessing the quality of an approximate function. Further, it will be demonstrated how the minimization of a suitable quantity derived from force may constitute a variational principle for bound states. The strategy applies also to Siegert states where traditional energy extremization principle ceases to work. Additionally, the utility of the force concept in semiclassical mechanics will be investigated.

*Corresponding author (e-mail: pchemkb@yahoo.com)

# 1. Introduction

In traditional quantum mechanics, the notion of force does not seem to play any vital role. Only, an average force is defined in the course of establishing a connection of quantum mechanics with Newton's laws of motion. This was done by Ehrenfest [1] who showed that

$$m\frac{d}{dt}\langle x\rangle_t = \langle p\rangle_t,$$
$$\frac{d}{dt}\langle p\rangle_t = \langle -\nabla V\rangle_t. \tag{1}$$

Here, the subscript *t* associated with the averages means that one is concerned with time-dependent states. This equation has another beauty. It shows not only that Newton's laws are obeyed in quantum mechanics *on the average*, but also provides a hint that classical dynamical variables are to be replaced by their *average* values in *time-dependent* states when one seeks a quantum-classical *correspondence*. In (1), $\langle -\nabla V\rangle$ is the force, or, more correctly, the *average* force. This force is, therefore, derivable from a potential and hence is conservative. However, there are other issues as well. First, the quantum-classical correspondence defined by (1) is not complete. This is because, if we define

$$F(x) = -\frac{\partial V}{\partial x}, \tag{2}$$

then, the association

$$\langle F(x)\rangle = F(\langle x\rangle) \tag{3}$$

holds only up to the harmonic oscillator case. If one goes beyond (*i.e.*, $V(x) = x^n$, or a linear combination of such, with $n > 2$), one has to bother about the dispersion in $x$ in (3). So, a relation like

$$m\frac{d^2}{dt^2}\langle x\rangle = F(\langle x\rangle), \tag{4}$$

equivalent to Newton's laws in a compact form, and more complete than (1) in respect of classicality, does not follow quite *generally* in the quantum domain. Secondly, for a time-independent state, the left hand sides in (1) vanish. Hence, the average momentum is zero, so is the average force. Therefore, $\langle F\rangle$ with a fixed known value of *zero*, appears useless for *stationary* states. This observation alone is sufficient to explain why the concept of force is not usually exploited in approximate calculations for energy eigenstates, barring exceptions like molecular force-field calculations that are 'classical' in nature.

It is, however, true that construction of approximate stationary states forms an active area of research in quantum chemistry. This is because, most of the problems of practical interest are not solvable exactly. On the other hand, chemical and spectroscopic properties are chiefly understood from such states only. Therefore, here we like to focus attention on how far this force concept can be made useful in a very general context of energy eigenvalue problems. In view of the fact that

$$\langle F\rangle = \langle\Psi|F|\Psi\rangle = 0, \Psi \neq \Psi(t), \langle\Psi|\Psi\rangle = 1, \tag{5}$$

as discussed just above, we shall concentrate on $F$ itself. The definition (2) is useful at this juncture because, like $V$, $F$ is a multiplicative operator. Basically, it's a function of coordinates (we shall not talk of non-conservative forces here). So, we can work with bare *F*. To obtain this quantity, we need to differentiate the Schrödinger stationary equation once. The process removes the energy, and, as we shall see, that adds to the advantage.

Our scheme of relegating the energy is significant also from a different standpoint. This concerns the quality of an approximate wave function and its assessment. In lieu of a detailed discussion, we may proceed via a few pertinent questions. Some of these are the following: When will a function $\varphi$ with an associated energy $\varepsilon$ be said to better represent a target state $\Psi$ than another function $\theta$ with energy $\acute{\varepsilon}$? Is it sufficient to inquire if $\varepsilon$ is closer to *E*? Is it even more comfortable if we have $\acute{\varepsilon} > \varepsilon > E$? Can some conclusion be drawn for ground states alone, or, at least? What about excited states, then? Does a definitive conclusion follow for variational calculations? What kind of variation is then allowed? What about other properties obtained from $\varphi$ and $\theta$? Well, such queries do not have specific answers. Indeed, they lose significance as soon as we realize that the goodness of a wave function should refer to *all* the properties of the state, *not* merely the energy. And, the standard wisdom is, the energy is mainly determined by the contributions from $\varphi$ around the potential minima. Therefore, the overall nature of a function is unlikely to be reflected through $\varepsilon$. Thus, several criteria of measuring the goodness of an approximate function like $\varphi$ or $\theta$ have emerged from time to time without involving $\varepsilon$ or $\acute{\varepsilon}$ straightforwardly. The most celebrated criterion of this sort, and perhaps the oldest, is due to Eckart [2]. He concentrated on the overlap $|\langle\varphi|\Psi\rangle|$ for the *ground* state, but obtained a bound to it that needed information about exact energies of both ground and first excited states. This is a limitation. A few others like (i) the local energy method [3], (ii) the least squares method (LSM) [4], (iii) measurement of time-stability [5] and (iv) satisfaction of specific hypervirial relations [6] use *H*, but not $\varepsilon$ directly. More recently, we employed [7] a recipe that rests on spatial derivative of the local energy. We hope that the present strategy, free from energy, can also lead to some such novel criterion.

Certain schemes of measuring the goodness of $\varphi$ additionally offer us suitable extremum principles to find an optimum $\bar{\varphi}$ (from a given trial function $\tilde{\varphi}$ containing one or more parameters) that satisfies *best* the target goodness property. In this connection, mention may be made of the LSM that, unlike the conventional linear variational method (LVM) [8], applies to both bound and resonant (Siegert) states. This naturally has prompted us to explore how the present endeavor performs in this regard too.

A different but related area that we like to cover is the semiclassical domain. In bound-state calculations, a Wilson-Sommerfeld (WS) type of approach [9] has been favorite for long. We shall investigate the role of force in this type of approximation and try to extract some useful information.

Our organization is as follows: In Section 2, we shall outline the scheme and point out its difference from a few others in vogue. This helps in delineating its role as an independent criterion when one proceeds to estimate the goodness of an approximate stationary state. Section 3 is intended to assess the vital status of force for exact and approximate stationary states. A particular concern on nodes may be found here. A few problematic situations involving inexact states will also be highlighted with explanations. The quality of approximate stationary states will be tested in some detail in Section 4. We choose a simple system and study it from a variety of angles here. A force-based variational scheme for both bound and Siegert states will be presented in Section 5. The performance of the scheme vis-à-vis a few other standard and recent schemes will be found here. Implementation of the study in the context of LVM is also possible. Section 6 will briefly concentrate on the importance of the concept in semiclassical WS type theories. Finally, we shall summarize the outcome of the whole endeavor in Section 7.

**2. The scheme**

The starting idea is very simple. We take the Schrödinger stationary equation and differentiate it once with respect to the positional coordinate. In 1-d, the relevant equations are

$$-\frac{\hbar^2}{2m}(\Psi''/\Psi) = (E-V), \qquad (6)$$

$$-\frac{\hbar^2}{2m}(\Psi''/\Psi)' = F. \qquad (7)$$

In going from (6) to (7), we get rid of the unknown energy, while the force shows up in a natural way. For some *n*-th stationary state, we have, instead of (7), the equation

$$-\frac{\hbar^2}{2m}(\Psi_n''/\Psi_n)' = F. \qquad (8)$$

The advantage over (6) is clear. We normally need to deal with two unknown quantities $\Psi$ and $E$ in (6), but only one is left in (7) or (8). Note that the removed energy is obtainable at any point as an eigenvalue of $H$, the Hamiltonian.

An approximate eigenstate of energy $\varphi$, be it ground or excited, does not satisfy (6). However, often we can find an effective Hamiltonian $H_0$ for which it is an eigenfunction. Denoting the potential part in $H_0$ by $V_0$, we then write

$$\left(-\frac{\hbar^2}{2m}\nabla^2 + V_0\right)\varphi = \varepsilon\varphi. \qquad (9)$$

From this energy eigenvalue equation, we do have a derived force $F_0$ defined by

$$-\frac{\hbar^2}{2m}(\varphi''/\varphi)' = F_0. \qquad (10)$$

Evidently, $F_0$ would differ from $F$, as long as $\varphi$ does not coincide with any eigenfunction of $H$. Here too, getting the energy (*not* $\varepsilon$ in this case) is no problem; it may be found as an average value, $<H>$. Further, it follows from (10) that, given any time-independent function $\varphi$, one can always define an $F_0$ based on this function. This is significant whenever we try to associate $\varphi$ with some energy eigenfunction.

Our next task will be to compare $F$ and $F_0$. When two approximate functions $\varphi$ and $\theta$ differ, their corresponding $F_0$ values will also be different. One expects that only when $\varphi \to \Psi$ (i.e., any $\Psi_n$), $F_0 \to F$, and vice versa. Here, the approach of $\varphi$ towards $\Psi$ should be taken in a norm sense, *i.e.*,

$$\|\varphi - \Psi\| \to 0. \qquad (11)$$

If $V_0$ and $V$ differ by a constant, $F_0 = F$, and hence $\varphi = \Psi$. The comparison of $F$ and $F_0$ can be made in various ways. But, our discussion in Sec. 1 indicates that both $<F>$ and $<F_0>$ would be individually zero. In this situation, one immediate bypass is to consider the ratio $F/F_0$, or its inverse, whichever is more convenient, and then take the *average*. Other possibilities of course exist (see later). However, sticking to the indicated ratio, we define

$$\mu = \langle \varphi | F/F_0 | \varphi \rangle, \mu(0) = 1, \langle \varphi | \varphi \rangle = 1. \tag{12}$$

Here, μ stands for the error in φ relative to its ideal value μ(0) = 1. Indeed, the error will be given by (μ-1) = μ′ and a smaller value of |μ′| is indicative of a better energy eigenfunction in respect of this force-based criterion. The recipe (12) is independent of state $n$. It is dimensionless as well. Further, by taking some normalized $\tilde{\varphi}_n$ containing one or a number of parameters embedded in it, one can construct $\tilde{\mu}_n$ following (12) and minimize its difference from the desired value of unity. Thus,

$$\min |\tilde{\mu}'_n| = \min |\tilde{\mu}_n - 1|$$

forms a variational principle.

It remains to be checked how far the present scheme differs from other related ones. One may be interested to also know if there exists some kind of kinship of the current strategy with the prevalent ones, although the genesis may be quite different. To this end, we now pay attention to a few very relevant methods. One popular measure of the error in φ is provided by

$$\Delta \varepsilon_n^2 = \langle \varphi_n | H^2 | \varphi_n \rangle - \langle \varphi_n | H | \varphi_n \rangle^2, \tag{13}$$

where we now specify that our chosen function φ is trying to represent some $n$-th state of Ψ, and hence the subscript. This quantity should be zero for an exact eigenstate of $H$. The dispersion (13) in energy has an extra advantage. In place of a fixed function $\varphi_n$ above, if we employ a trial wave function $\tilde{\varphi}_n$ containing certain parameters, the minimization of $\Delta \tilde{\varepsilon}_n^2$ with respect to such parameters leads to an independent variational scheme. This precisely is the LSM. However, one must admit that (13) is a more direct measure of the error in *energy*, not of φ.

Next, we start from the definition of the local energy in one dimension, given by

$$e_n(x) = \frac{H \varphi_n}{\varphi_n}.$$

In view of wide variations of $e_n(x)$ over the entire space, several alternatives have emerged from time to time. Here, we consider a recent work [7] based on it. For an exact state, $e_n(x)$ would be a constant; hence $e_n'(x)$ would vanish, where the prime refers to differentiation with respect to $x$. Therefore, it seems natural to estimate the quantity $\langle e_n'(x) \rangle$ as a measure of error in $\varphi_n$. However, this integral may not reflect the true state of affairs because of partial cancellation of positive and negative contributions [note that $e_n'(x) \geq 0$ is not ensured]. The situation becomes worse when $H(x) = H(-x)$. The eigenstates are then either even or odd. In such a case, $\langle e_n'(x) \rangle = 0$ by symmetry alone. Therefore, we defined

$$\eta_n = \langle \varphi_n | [e_n'(x)]^2 | \varphi_n \rangle \tag{14}$$

and advocated use of the quantity $\eta_n$ to measure the error in $\varphi_n$. We do not require *any* information about exact energy eigenstates in this definition. It is apparent from (14) that one may choose a trial function and try to minimize the error

$$\tilde{\eta}_n = \langle \tilde{\varphi}_n | [\tilde{e}_n'(x)]^2 | \tilde{\varphi}_n \rangle, \tag{15}$$

thus obtaining another variational route to fix the function $\tilde{\varphi}_n$. This is essentially a kind of *error minimization* scheme.

Let us note here a few points. If $\varphi_n$ satisfies the equation $H_0 \varphi_n = \varepsilon_n \varphi_n$, where $H_0 + \upsilon = H$, then $e_n'(x) = \upsilon'$. Thus, $\eta_n$ in (14) measures the *mean square excess force*. For a constant $\upsilon$, $\eta_n$ becomes zero, implying that the wavefunction needs no modification. In the LSM, one identifies $\Delta \upsilon^2$ with the error (13). Here too, $\Delta \upsilon^2$ becomes zero when $\upsilon$ is constant. Otherwise, a smaller value of $\Delta \upsilon^2$ only renders $<\upsilon>$ more meaningful, but the latter does not reduce in magnitude. Therefore, the connection of $\Delta \upsilon^2$ with $\upsilon'$ is not direct. On the contrary, criterion (14) measures $<(\upsilon')^2>_n$. It relies straight on the *rate* of change of *excess potential* $\upsilon$. More direct, however, is the present approach (12) where the *average relative force* is measured.

In case of practical systems, the parameters embedded in $\tilde{\varphi}_n$ may be quite large in number. Then, optimization may be done, for example, via *random variations* [10]. For the ground state, the procedure is straightforward. Particularly, it is simpler for nonlinear variations. In case of excited states, one needs some extra constraints involving orthogonality and decoupling integrals with approximate lower states during the optimization process. This is a common feature of nonlinear variations. Therefore, excited bound states are better treated through LVM to obtain some $\bar{\varphi}_n$. One may subsequently employ (12), or any such other measure, only to check its quality. Thus, to keep things simple and to highlight the workability of the present strategy, we shall be concerned here with ground bound states only. Further, in view of the success of the LSM, here we are tempted to explore how far the satisfaction of (12) by choosing a trial $\tilde{\varphi}$ can be of value in studies on Siegert states as well. While resonant eigenfunctions are not truly square-integrable, we know that normalizable functions can yield real energy parts ($E_r$) for such states. Indeed, the LSM has been shown [11] to offer $E_r$ rather directly. For convenience, however, we shall henceforth drop the subscript *r* in $E_r$ as the procedures that we follow in this work yield real energies only.

Apart from the analytical studies, here a few model systems will be chosen for numerical demonstrations on which results via a variety of procedures are either available or can be estimated without much trouble. The advantage is, such data are free from additional approximations that are often invoked in tackling more real-life problems. Hence, any assessment made on the basis of these calculations will be free from undesirable errors.

### 3. Force and energy eigenstates
*3.1. Stationary states*
We first consider bound, stationary states that are exact. We then have from (7)

$$\langle F \rangle = -\frac{\hbar^2}{2m} \int_{-\infty}^{\infty} |\Psi|^2 \left(\Psi''/\Psi\right)' dx$$
$$= -\frac{\hbar^2}{2m} \left[(\Psi')^2\right]_{-\infty}^{\infty} = 0 \quad (16)$$

for a real function in 1-dimension. This is an alternative route to arrive at the result that $<F>$ is zero, as expected [*cf.* the discussion below (4)]. Note that nodes in $\Psi$ do not cause any problem here as $(\Psi''/\Psi)$ does not possess any singularity. One can use (7) or (8) to obtain bare $F$ as well. Indeed, while $<F>$ does not contain any more information, bare $F$ does. As preliminary examples, consider a few situations: (i) For a free particle with wave function exp[$ikx$], it is easy to check that $F = 0$. (ii) Eigenfunctions of energy for the particle-in-a-box also show $F = 0$ for any state. (iii) The 3$^{rd}$ excited stationary state $x(3 - 2x^2)$exp[-$x^2$/2] for a harmonic oscillator also leads smoothly to $F = -(\hbar^2/2m)2x$, which is exact and is same as that derived from any other state. But, this is not all. Nodes are important in this context because of a specific connection with force, to be seen below.

To continue, we start with a useful ansatz [12, 13] for a general bound state $\Psi_n$. This is given by
$$\Psi_n = f_n \exp[g_n]. \tag{17}$$
Here, the first factor $f_n$ is a polynomial that accounts for nodes while the second one ($g_n$) ensures an exponential fall-off of the wave function that is associated with most potentials supporting bound states. The choice (17) yields, suppressing temporarily the subscript $n$,
$$-\frac{2m}{\hbar^2}F = \left(\frac{f''+2g'f'}{f}\right)' + \left(g''+(g')^2\right)'. \tag{18}$$
Two remarks are now in order. Depending on the nature of the potential, one can examine, without solving the problem, whether the exponential part in (17) will be the same for any state $n$. This part itself is interesting. One needs to employ a suitable trial function of the form
$$\Psi_n \sim x^{n-1}\exp[-\beta_n|x|^\delta], \tag{19}$$
in case of 1-d oscillators with potential $x^{2N}$, for example, and study the large-$x$ behavior. An immediate finding is that both $\beta_n$ and $\delta$ are independent of $n$ (indeed, $\beta_n \sim 1/\delta$, $\delta = N + 1$). Hence, the exponent $g_n$ in (17) does *not* depend on $n$ for such problems. One can further check that, while for a bare H-atom the large-$r$ behavior does show an $n$-dependence of the exponent $g_n$, a radial perturbation $\lambda r^M$ on the same system shows again an independence. However, more interesting now is the following outcome of (18). Once we are sure that $g_n$ is the same for any $n$, we see that the second part within parentheses of (18) accounts for the total $F$, since for the ground state we can take $f = 1$. This implies, the first part at the right side of (18) should not contribute anything for excited states. The message is clear. One must then have the condition
$$f_n''+2g_n'f_n' = c_n f_n \tag{20}$$
where $c_n$ is a constant for a specific state. Thus, the polynomial $f_n$ that contains the nodal information must satisfy (20). This is a nice result in a compact form. One can apply (20) to generate, *e.g.*, the Hermite polynomials, taking $g' = -x$ for the $x^2$ potential case.

*3.2. Approximate stationary states*

For an approximate stationary state, one arrives easily at an equation equivalent to (16) yielding the average value of $F_0$:
$$\langle F_0 \rangle = -\frac{\hbar^2}{2m}\int_{-\infty}^{\infty}|\varphi|^2(\varphi''/\varphi)'dx$$
$$= -\frac{\hbar^2}{2m}\left[(\varphi')^2\right]_{-\infty}^{\infty} = 0. \tag{21}$$

But, more useful is the local force $F_0$ with which one can compare the true $F$ for the given problem. We choose now a few situations to examine how such a comparison helps. Here and henceforth we shall take $\hbar = 1$ and $m = \frac{1}{2}$.

First, take the particle-in-a-box case. We employ the function $x(L - x)$, with $0 \leq x \leq L$, that is known to furnish a good quality energy as an approximation to the ground state. However, this function yields

$$F_0 = -\frac{2(L-2x)}{x^2(L-x)^2} . \tag{22}$$

Actually, $F$ is zero at any point $x$ within $0 \leq x \leq L$. From (22), we can see that $F_0 = 0$ only at $x = L/2$. This observation offers a good lesson. Around the maximum probability region, we have $F_0 \approx F$ and this region is chiefly responsible for the goodness of energy. Hence, although (22) does not mimic the behavior of the true $F$, the chosen function gives surprisingly good energy. One also notes from (22) that, near the boundaries, *i.e.*, at $x \approx 0$ and $x \approx L$, we get $F_0 \to -\infty$. The departure from exactness is thus most pronounced around these regions. While the energy is not affected due to very low probabilities, $\Psi_0'$ really shows large departures from $\varphi'$ when $x \approx 0$ and $x \approx L$. Thus, goodness of energy does not mean, in anyway, the goodness of a given function as an eigenfunction. A similar series of conclusions follows for higher approximate states, *e.g.*, the function $x(L - x)(L/2 - x)$ approximating $\Psi_1$, etc.

Secondly, we take up the harmonic oscillator case given by the Hamiltonian

$$H = -\frac{d^2}{dx^2} + \lambda x^2 . \tag{23}$$

As a trial function for the ground state, we choose

$$\widetilde{\varphi} = A^2 - x^2, -A \leq x \leq A \tag{24}$$

and zero otherwise. This function also delivers good quality energy after a due optimization with respect to the parameter $A$. A simple calculation, however, shows

$$F_0 = \frac{4x}{\left(A^2 - x^2\right)^2}. \tag{25}$$

This may be contrasted with the actual force $F = -2\lambda x$. We note that $F$ and $F_0$ are of opposite signs! But, once again, we do see also that $F_0 \approx F$ around $x = 0$, the region primarily responsible for the goodness of energy. Next, we notice that the function (24) gives

$$\widetilde{\mu}_0 = \langle F/F_0 \rangle = -\frac{8\lambda}{21} A^4, \tag{26}$$

which follows from (12). For bound states, we need to have $\lambda > 0$ and hence the minimization scheme, discussed below (12) does not apply. It leads to an absurdity ($A = 0$). A standard bypass in such situations is provided by a *symmetric average*. We should here try instead

$$\min \left| \tfrac{1}{2} \langle (F/F_0) + (F_0/F) \rangle - 1 \right|.$$

In discussions around (12), we indicated about other possibilities of implementing the basic idea. What we quoted above is one such variant. This leads to minimization of the quantity

$$\frac{15}{8\lambda A^4} + \frac{4\lambda A^4}{21} + 1. \tag{27}$$

From (27), we obtain a relation between $A$ and $\lambda$ as $A \approx 1.33/\lambda^{1/4}$. It is remarkable that minimization of the average energy with respect to the trial state (24) yields a very similar form of variation, viz. $A = 2.05/\lambda^{1/4}$. The numerical factors differ significantly, however. One point behind the departure is that, at large distances from the origin, $F_0$ and $F$ differ widely. But, the mean energy can be lowered by allowing an enhanced delocalization. Thus, the force-based approach offers a lower estimate of optimized $A$ than the energy-based one.

Thirdly, we consider another simple situation. For the same $H$ in (23), we can use a different trial ground state function

$$\widetilde{\varphi} = \cos\left(\frac{\pi x}{2A}\right), \quad -A \leq x \leq A, \tag{28}$$

and zero otherwise. Like (24), this is also a good function in terms of energy. However, (28) yields $F_0 = 0$, as it refers exactly to the ground state of a particle in a box. For such functions, our force-based recipe (12) does not work. This is, of course, a weakness of the present endeavor, though we realize that the box problem is basically an idealization. In terms of potential also, a smooth transition from a box to an oscillator problem is impossible. Hence, the failure is not unnatural.

Having understood the pros and cons of applicability of (12) to diverse situations, we now come to the fourth and final important point. This concerns the extraction of nodal information in course of approximate calculations for excited states. Consider a problem that is not exactly solvable, *e.g.*, an anharmonic oscillator. Suppose, we know somehow a very good ground state. We also know from an asymptotic analysis that the exponent $g_n$ in (17) is independent of $n$. In such a case, we can immediately employ (20) to estimate the coefficients of the polynomial symbolized by $f_n$. This implies, in other words, that we have an *a priori* knowledge of the nodal positions. Let us emphasize here that a mere knowledge of the large-$x$ behavior of $g_n$ does not help; one has to know the *detailed* behavior to get the nodal information. However, once the nodal positions are known nearly exactly, we need not bother about the orthogonality and related constraints in the course of carrying out a nonlinear variation. To state otherwise, we can treat the problem as *constraint-free*. The exact nodes really take care of those constraints automatically. This is surely a big advantage from a practical point of view. Although slight errors in locating the nodal positions might inhibit the *upper boundedness* property of energy, very good approximate functions may thus be obtained. Thus, a nonlinear variation for excited states may be pursued freely.

**4. Testing the quality of approximate states**

Before testing the quality of some $\overline{\varphi}$ as an eigenfunction of a given Hamiltonian, it seems imperative to study first the sensitivity of the various existing criteria vis-à-vis the one being put forward here. A good indicator of the quality of an approximate eigenfunction must be sufficiently sensitive to small changes in parameter values, embedded in the function of choice.

To achieve the above end, we first choose the Hamiltonian in (23) at $\lambda = 1$. This is an exactly solvable problem. We choose here

$$\widetilde{\varphi} = N \exp[-\beta x^2], \tag{29}$$

where $N$ stands symbolically for the normalization constant. For any $\beta \neq \frac{1}{2}$, this function can be taken to represent an approximate eigenenergy function. We then proceed to estimate the following quantities: (i) $\mu_0'$, (ii) $\Delta\varepsilon_0^2$, (iii) $\eta_0$ and (iv) $<H>$. In Figure 1, we display the relevant variations as functions of $\beta$. Curve 1 shows the variation of $\mu_0'$, curves 2 and 3 show the same of $\Delta\varepsilon_0^2$ and $\eta_0$, respectively. Curve 4, for convenience, depicts the change of ($<H>$-1). We know that the exact state yields the eigenenergy value of unity at $\beta = \frac{1}{2}$. The figure clearly shows that the variation is least marked for the energy itself (curve 4). This clearly reveals the need of some other better criterion. Curve 2 shows a shade better sensitivity than 4, thus justifying the LSM. Curve 3 exhibits a *much better* variation around the optimum value of $\beta$. This refers to measure of the mean square extra force. It is comforting to note that the present endeavor too yields a nice, very sensitive dependence on $\beta$, as shown by curve 1 in the figure. It is additionally advantageous because of its passage from positive to negative values. A *sign change* is always welcome in respect of sensitivity. We also note that all such measures reveal that $\beta = \frac{1}{2}$ is the best choice. Figure 2 shows a similar sensitive dependence of $\mu_1'$ (curve 1) and $\eta_1$ (curve 3) relative to spread $\Delta\varepsilon_1^2$ (curve 2) and average energy (curve 4) variations [actually ($<H>$-3), for convenience] for the first excited state of the same system. The results displayed here correspond to the function $x\tilde{\varphi}$, with the same $\tilde{\varphi}$ as in (29). Again, the optimum $\beta$ is $\beta = \frac{1}{2}$ at which all the criteria are best satisfied. From these plots, we can definitely be convinced also about the necessity of introducing $\mu'$ as a new criterion.

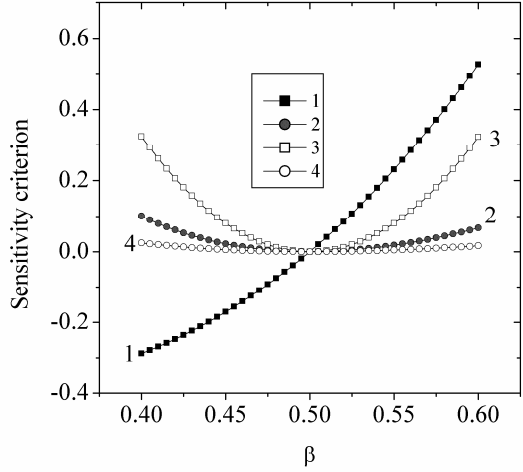

**Figure 1.** Behavior of different sensitivity criteria (see text for details) for the ground state of the harmonic oscillator as a function of the exponent $\beta$. The exact ground state corresponds to $\beta = 1/2$.

Next, we come to quality. To this end, we select the quartic anharmonic oscillator Hamiltonian defined by

$$H = -\frac{d^2}{dx^2} + x^2 + \lambda x^4. \tag{30}$$

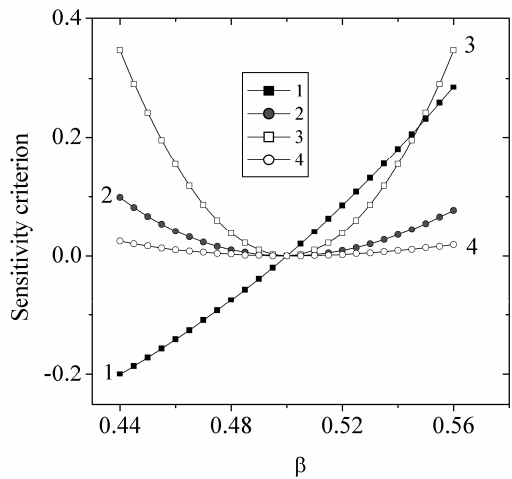

**Figure 2.** The same plot as in Figure 1, but now for the first excited state of the harmonic oscillator. Here too, $\beta = 1/2$ refers to the exact state.

Choosing the trial function (29), we can optimize the energy. It is likely that we can get a good ground state at small $\lambda$. Figure 3 shows plots of the relevant quantities of interest as functions of $\beta$. The dependence of $(\langle H \rangle - 1)$, as shown by curve 4, is indeed the weakest.

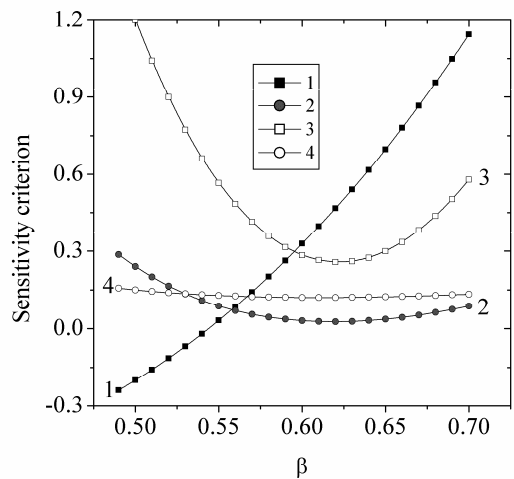

**Figure 3.** Behavior of different sensitivity criteria around the optimum $\beta$ ($\beta = 0.61$) yielding minimum energy for the ground state of the quartic anharmonic oscillator at $\lambda = 0.2$.

There is a minimum at about $\beta = 0.61$, but it's quite shallow. Thus, slight changes of $\beta$ do not affect the computed energy significantly. The spread in energy, depicted by 2 in the figure, shows also a small value around β = 0.61. One expects normally that the function (29) at this β is sufficiently good as an approximate ground state. A look at curves 1 and 3, however, tarnishes the myth. Curve 1 shows a significant positive error for $\mu_0'$ that could reduce at lower β. On the other hand, curve 3 indicates that a somewhat higher β might reduce $\eta_0$. These latter two criteria really reveal that a sizeable amount of error is still contained in the chosen function.

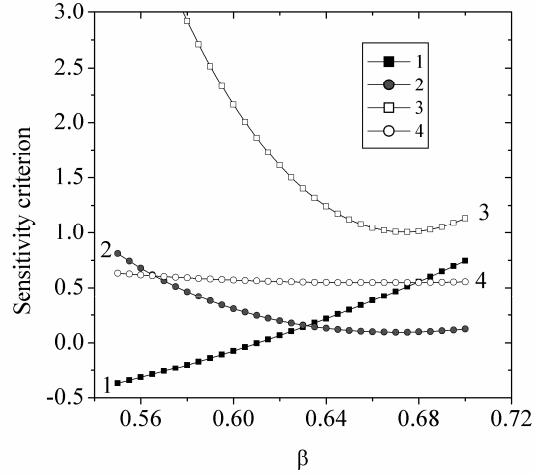

**Figure 4.** Same as Figure 3, but now around the optimum β (β = 0.66) yielding minimum energy for the first excited state of the quartic anharmonic oscillator at λ = 0.2.

Figure 4 shows similar variations of the concerned quantities for the trial state $x\tilde{\varphi}$, now intended to get information about optimized first excited state of (30). Here, curve 4 represents ($<H>$-3) that attains a minimum around β = 0.66, and we again note features akin to the earlier case that need not be reiterated. What is common to both the figures is that, our present criterion μ′ requires β < β(opt) while our earlier one (η) needs β > β(opt) for betterment. So, for gradually better wave functions, one should obtain μ′ closer to zero and η closer to its minimum simultaneously (as found in Figure 1 and Figure 2). This leads to a lesson. Since the above two criteria cannot both be satisfied at the same time with (29), we conclude that the choice of the trial function itself is *bad* (though the energy, one can check, is obtained fairly accurately). Such a conclusion could not follow so *emphatically* from a mere study of spread. The spread calculation is relatively involved too, unless simple systems are considered. Lastly, one does *not* require *any* accurate experimental or theoretical data to land at the inference. This surely serves as a major advantage of the approach.

The quality of Siegert states (see later for more detailed exposition) may be judged likewise. Suppose we take $H$ in (30) at λ = -0.2 and employ the trial state (29) to minimize the spread. During the minimization process, properties μ′, $\Delta\varepsilon^2$, η and $<H>$ change with β. Such changes are shown in Figure 5, respectively by curves 1-4. Note that here $<H>$ does not show any minimum with respect to variations of β. So, in the absence of other information, one is inclined to accept as β(opt) the value at which $\Delta\varepsilon^2$ is minimum. This occurs near β = 0.462 and gives a value of 0.8277 as the average energy. Here, however, we notice that μ′ is minimum around β = 0.439 where the average energy is 0.8142, while η is minimum around β = 0.464 where $<H>$ = 0.8286. Thus, one clearly finds that the LSM does *not* always offer a balanced state with respect to all the properties. Our discussion in the subsequent section will make the point clearer.

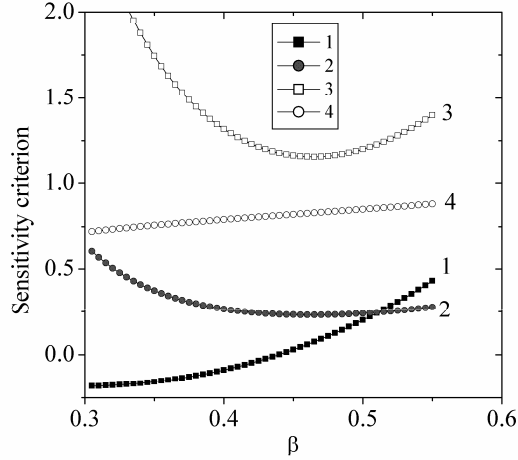

**Figure 5.** Plot of various sensitivity criteria (see text) vs. β for a Siegert state at λ = -0.2 of H in (30) using the trial function (29).

**5. Force-based variational calculations**
*5.1. Bound states*

We now consider the possibility of employing $\tilde{\mu}'$ as a variational functional that may be subsequently optimized, and check whether it can offer a good, approximate eigenfunction of energy. Additionally, we shall go for optimizations of $\Delta\tilde{\varepsilon}^2$ and $\tilde{\eta}$ for comparative purposes, along with minimization of the average energy. Let us call such schemes respectively as Scheme I, Scheme II, Scheme III and Scheme IV.

To continue, we take the form (29) for $\tilde{\varphi}$ and the Hamiltonian (30), and display the results in Table 1, at λ = 1, for the ground state. For the first excited state, as before, we choose $x\tilde{\varphi}$ as the function. The corresponding results are also shown in the same table. For convenience, we quote now some near-exact data, based on earlier calculations [15]. While these are not necessary, they can additionally guide us towards proper conclusions. The ground-state energy is 1.39235 where the kinetic part contributes an amount $<T>$ = 0.8263. The mean square displacement is $<x^2>$ = 0.3058. From (29), one finds that $<T>$ = β, while $<x^2>$ = 1/(4β). The first excited state lies at energy 4.6488 with the kinetic part contributing the amount 2.8321. For this state, $<x^2>$ = 0.8013. The function $x\tilde{\varphi}$ here yields $<T>$ = 3β and $<x^2>$ = 3/(4β). With these results in mind, the following few points are notable: (i) Scheme I always gives a lower estimate of $<T>$; thus, it favors an enhanced delocalization. (ii) Scheme III, on the contrary, always overestimates the kinetic part. (iii) Scheme II occupies an intermediate position, so does Scheme IV, but the former leaning a bit more towards Scheme III. (iv) Scheme I and Scheme III show much wider variations than Scheme II. (v) Scheme IV exhibits the least variation over the range of β considered here. The particular nature of such variations of the quantities is also pretty clear from figures 3 and 4, though in a less pronounced manner because of lower λ. The significantly different data obtained via optimizations of the four schemes considered here point only to a bad choice of the trial function. A better function with more embedded parameters would have shown much closer results. This did occur in figures 1 and 2.

| State | Scheme | β | μ′ | $\Delta E^2$ | η | E(avg) |
|---|---|---|---|---|---|---|
| 0 | I | 0.662 | 0 | 1.136 | 8.600 | 1.467 |
|   | II | 0.866 | 2.464 | 0.179 | 2.392 | 1.405 |
|   | III | 0.878 | 2.665 | 0.181 | 2.376 | 1.406 |
|   | IV | 0.836 | 1.999 | 0.192 | 2.569 | 1.403 |
| 1 | I | 0.836 | 0 | 1.728 | 17.975 | 4.747 |
|   | II | 0.977 | 2.507 | 0.541 | 8.249 | 4.681 |
|   | III | 0.988 | 2.739 | 0.546 | 8.202 | 4.684 |
|   | IV | 0.952 | 1.998 | 0.571 | 8.693 | 4.678 |

**Table 1.** Results of different schemes (see text) of optimization of β in $\widetilde{\varphi}$ and $x\widetilde{\varphi}$ [see (29)] for the ground and first excited states, respectively, of the system (30) at λ = 1.

*5.2. Siegert states*

The case of Siegert states [14] is different. These are *metastable* quantum states and hence are significant in a variety of contexts. Here, the system Hamiltonian does *not* support any bound state. The shape of the potential is such that *at least* one local minimum exists, but there is also the provision of any bound state in this minimum to *tunnel* out (shape resonances). Therefore, the standard methods for eigenenergy functions calculations, e. g. the variational method, do not apply. Indeed, if we prepare a square-integrable packet centred at a metastable minimum, it will evolve with time as the state is not strictly stationary. However, a continuous spectrum of *H* forbids the state from pursuing *recurrences*. Instead, the state decays. Therefore, question of stability of the packet arises. Primarily, these resonant states have attracted attention over the years because of methodological interest. Certain response properties (e. g., the polarizability) also need specific external perturbations that make the overall system Hamiltonians yield such Siegert states.

When we concentrate on Siegert states, we notice that only some *specific* properties become important. The *lifetime* is one such property that reflects the time-stability. One way to estimate it is via the energy spread $\Delta E_n^2$ that measures the short-time stability. This is the usual route in case we take *real* functions to describe such a state. An alternative is to choose a *complex* function for the state that yields a complex energy. The imaginary part of this energy is linked with the lifetime. The spatial stability is guided by the real part of energy, or <H> in the former approach. Another property of concern is the *localization*. It is usual to measure it through <$x^2$> (actually it's square root) because in many situations <x> could be zero for reasons of symmetry alone.

To study the efficacy of $\mu'$ in calculations of Siegert states, we again take $H$ in (30), but now with $\lambda < 0$ that allows shape resonances. Using the trial function (29) and requiring that $\tilde{\mu}'_0 = 0$, the best possibility, one obtains

$$8\beta^3 - 2\beta - \lambda = 0. \tag{31}$$

Real solutions for β from this equation at varying λ-values are shown in Figure 6. The acceptable (β > 0) solution at λ = 0 is represented by point A in the curve. As λ is varied towards the positive side, the solution follows the curve AB. No problem is encountered along this side. But, if we choose the λ < 0 region, *two* acceptable solutions appear. The solution that is an analytic continuation of the λ = 0 point should then be taken. But, this proceeds up to point C along AC, at which there is a *crossover* to the other region. Therefore, it turns out that such metastable states can exist up to the point C, *i.e.* up to λ = -0.3849, to be precise. This is an interesting point by itself as *no* such *limit* is provided by the LSM, though we are sure that a larger negative λ reduces the lifetime and hence, beyond a point, the notion of an 'almost'-bound stationary state should break down. One confronts a virtually similar situation with the trial state $x\tilde{\varphi}$ for which the condition (31) is replaced by

$$8\beta^3 - 2\beta - 3\lambda = 0, \tag{32}$$

which follows from $\tilde{\mu}_1' = 0$. The β-λ plot in this case is very much like Figure 6; only, the point C here would refer to λ = -0.1283, one-third of the earlier limit. We may mention that 'excited' Siegert states (note the node at $x = 0$) do not seem to have been studied before. But, it is natural to expect a smaller negative λ limit for excited states because of increased spread of the wave function. In other words, the spatial localization of a state decreases as it becomes more excited. In both the cases, point C corresponds to β ≈ 0.29. However, we also notice that while $<x^2>$ = 1/(4β) for the nodeless state, the value changes to 3/(4β) for the other state. Hence, one comes to the conclusion that, roughly 75% flattening of the function relative to the unperturbed value of $<x^2>$ is allowed in both the situations. If the function stretches out more, probably it loses the desired localization property that characterizes the state as a Siegert state.

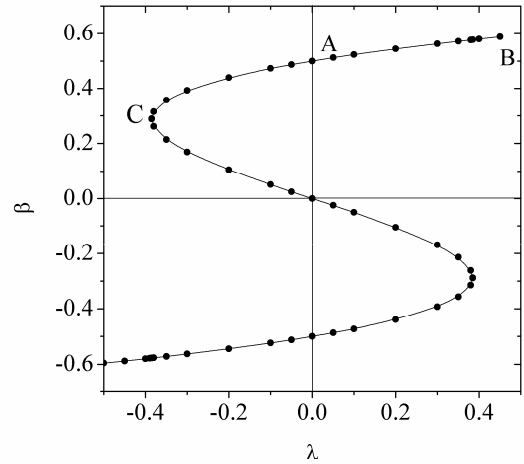

**Figure 6.** Plot of ß vs. λ that shows the acceptable region at negative λ, according to (31).

Table 2 shows the energies of Siegert states of the Hamiltonian $H(\lambda) = -\nabla^2 + x^2/4 + \lambda x^4/4$ obtained by adopting several schemes. For comparison purpose, one bound-state

| λ | Scheme I | *Scheme II | Scheme III | *E(A) | #E([2/1]) | *E(Num) | $E(S) |
|---|---|---|---|---|---|---|---|
| +0.05 | 0.5348 | 0.5331 | 0.5331 | 0.5332 | 0.5328 | 0.5327 | 0.5327 |
| -0.01 | 0.4924 | 0.4923 | 0.4923 | 0.4922 | 0.4922 | 0.4922 | 0.4922 |
| -0.03 | 0.4763 | 0.4750 | 0.4751 | 0.4742 | 0.4744 | 0.4742 | 0.4742 |
| -0.05 | 0.4588 | 0.4548 | 0.4554 | 0.4507 | 0.4516 | 0.4507 | 0.4512 |

*From Ref. 16.      #From Ref. 17.      $From Ref. 18.

**Table 2.** Comparative energies of Siegert states of $H(\lambda) = -\nabla^2 + x^2/4 + \lambda x^4/4$ via various methods.

result (λ > 0) is displayed as well. Apart from the three schemes under survey here, a number of values from other sources are available. In the table, $E(A)$ refers to the sum of the perturbation series for ground-state $E(\lambda)$ up to the numerically smallest term, called the asymptotic sum. Along with numerical and LSM results, these $E(A)$ values are found in Ref. 16. A useful Padé approximant [17] to the parent $E(\lambda)$ series yields $E([2/1])$. The $E(S)$ values are found by adopting the stabilization method, considered as another efficient method of studying resonances (see Ref. 18 for details). We may mention also that a specific reference to the trial function (29) is made only in results of schemes I - III, not in others. The table does not, however, reveal anything special with Scheme I. But, there is one more point that deserves notice and there we shall see how Scheme I is favored. To this end, we refer to Figure 7 where optimum β values, as obtained from schemes 1-3, are plotted at different negative λ for $H$ in (30). The figure shows a monotonic variation of β in Scheme I (see curve 1). The other two schemes exhibit passage through minima. As a result, at smaller negative λ, the state concerned is more localized in Scheme I. The other two schemes yield states with rapidly increasing $<x^2>$, and the LSM (Scheme II) performs worst in this respect.

In Table 3, we present a comparative account of the performance of various schemes in the same spirit as of Table 1. A well-balanced state is lacking, except the one obtained around λ ≈ -0.175. However, this can be misleading too, because one should doubt the adequacy of Scheme II and Scheme III beyond the point of their respective minima observed in Figure 7 that occurred roughly around $\lambda \approx -0.125$. Thus, Scheme I can sometimes really help in ascertaining the 'goodness' of other measures too. Further, if we accept that the observed minima have something to do with the applicability of the schemes, Scheme I is better in this respect as well, because it acts over a wider range, up to $\lambda \approx -0.385$.

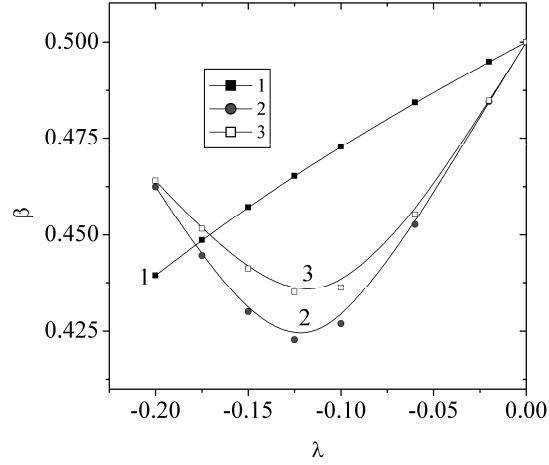

**Figure 7.** Plot of optimized ß at different negative $\lambda$ values for the three schemes.

*5.3. Case of linear variations*

So far, we have been concerned with nonlinear variations. But, the measure $|\mu'|$ in terms of force applies to the LVM with equal facility. Briefly, here one has to proceed as follows: Suppose we start with an orthonormal set of functions $\{\phi_j\}$ that satisfies
$$H_0 \phi_j = \varepsilon_j \phi_j . \tag{33}$$
Given the Hamiltonian $H$ for which energy eigenstates $\Psi$ are sought, we write
$$\Psi = \sum_{k=1}^{N} c_k \phi_k . \tag{34}$$
The coefficients $c_k$ in (34) are evaluated by using the standard method for *each* discrete energy state found after diagonalization of the concerned Hamiltonian matrix. We assume that this has already been done. If we further write that $H_0 = T + V_0$ and $H = H_0 + \upsilon$ where $T$ is the kinetic energy operator, a simple manipulation leads to the equation
$$F_0 = \frac{\sum_{k=1}^{N} c_k \varepsilon_k \phi'_k}{\sum_{k=1}^{N} c_k \phi_k} - \frac{\sum_{k=1}^{N} c_k \varepsilon_k \phi_k \sum_{j=1}^{N} c_j \phi'_j}{\left(\sum_{k=1}^{N} c_k \phi_k\right)^2} - V'_0 . \tag{35}$$

The above expression refers to the approximate force. The actual force should be
$$F = -\upsilon' - V_0'. \tag{36}$$

| λ | Scheme | β | μ' | ΔE² | η |
|---|---|---|---|---|---|
| -0.02 | I | 0.495 | 0 | 0.002 | 0.008 |
|  | II | 0.485 | -0.039 | 0.001 | 0.005 |
|  | III | 0.485 | -0.038 | 0.001 | 0.005 |
| -0.06 | I | 0.484 | 0 | 0.014 | 0.079 |
|  | II | 0.453 | -0.103 | 0.008 | 0.059 |
|  | III | 0.455 | -0.096 | 0.008 | 0.059 |
| -0.1 | I | 0.473 | 0 | 0.044 | 0.237 |
|  | II | 0.427 | -0.131 | 0.033 | 0.208 |
|  | III | 0.436 | -0.108 | 0.033 | 0.206 |
| -0.15 | I | 0.457 | 0 | 0.113 | 0.589 |
|  | II | 0.430 | -0.074 | 0.108 | 0.585 |
|  | III | 0.441 | -0.045 | 0.109 | 0.582 |
| -0.175 | I | 0.449 | 0 | 0.165 | 0.848 |
|  | II | 0.445 | -0.011 | 0.165 | 0.849 |
|  | III | 0.452 | -0.009 | 0.166 | 0.848 |
| -0.2 | I | 0.439 | 0 | 0.235 | 1.179 |
|  | II | 0.462 | 0.066 | 0.231 | 1.156 |
|  | III | 0.464 | 0.071 | 0.231 | 1.156 |

**Table 3.** Adequacy of different schemes (see text) of optimization of β in $\tilde{\varphi}$ [see (29)] for the nodeless resonant state of the system (30) at various negative λ-values.

It is now easy to get |μ'| and see whether it tends to vanish, and, if so, how. Particularly, such a scheme is likely to work very well in problems of studying *basis saturation*. Indeed, one rarely checks any property other than the convergence of energy with $N$ for a given state (usually ground) to ensure virtual completeness of basis. But, this force-based criterion has already been seen to be much more sensitive and hence can guide us towards the right choice of $N$ for a given problem. Pilot calculations in this context may be pursued to gain more insight.

**6. Force in a semiclassical context**

Another interesting application of the force concept may involve the semiclassical domain. For convenience, here we first cast the Schrödinger energy eigenvalue equation
$$-\frac{\hbar^2}{2m}(d^2\Psi/dx^2) = (E-V)\Psi \tag{37}$$
in the form of a Riccati equation:

$$-\frac{\hbar^2}{2m}(\chi^2+\chi')=(E-V). \tag{38}$$

Here, $\chi$ stands for $d(\ln\Psi)/dx$. In WS formulation, one neglects the second term in the parenthesis at the left side. Thus, in place of the true force

$$F=-\frac{\hbar^2}{2m}(2\chi\chi'+\chi''), \tag{39}$$

one actually takes here a 'semiclassical' force

$$F_{SC}=-\frac{\hbar^2}{m}\chi\chi'. \tag{40}$$

The conclusion is also apparent from the WS choice

$$\chi=ip/\hbar \tag{41}$$

with $p$ as the momentum. Indeed, putting ansatz (41) in (39), one sees that the neglected part in (40) corresponds to the $(d^2p/dx^2)$ type of term. We know, such a term is disregarded in WS or related theories. One may now inquire how far (40) is going to be useful. Specifically, we like to examine whether $F_{SC} \to F_0$ in the large-$n$ (Bohr) limit. This might have established the Bohr correspondence principle from a different angle. Note that (40) is exact for the free particle and harmonic oscillator (ground state) problems. In both these cases, $\chi'' = 0$. However, if we employ it to the particle-in-a-box case in $(0, L)$, another exactly solvable problem, it turns out that

$$\chi''=-2\chi\chi'=2(n\pi/L)^3\cot\frac{n\pi x}{L}\cosec^2\frac{n\pi x}{L}. \tag{42}$$

This result is obtained by using the exact energy eigenfunction. The $n$-dependence does not help to make the right side vanishingly small. From such a counter-example, we conclude that

$$\lim_{n\to\infty}F_{SC} \not\to F. \tag{43}$$

This means, from a study of exact states, we do not have $F_{SC} \to F$. Therefore, there exists some other mechanism through which these semiclassical results match the true ones.

Another point of concern is the appearance of nodes. Normally, we are aware that the WS type theories work nicely at large $n$. But, the probability distributions found subsequently do not possess $n$ nodes. Why? An answer is provided by (40). Let us consider the harmonic oscillator as a test case. Choose the potential as $x^2$. Then $F = -2x$. Neglecting the second term in (39) in going for WS type schemes, we put it straight in (40) to get

$$\chi\chi'=x \tag{44}$$

with $\hbar = 1$, $m = \frac{1}{2}$. From this equation, one gets a solution for $\chi$, and hence of $\Psi$, that does not contain any nodes. Indeed, if we impose on $\Psi$ the form (17) to take into consideration the nodal pattern through $f$, we obtain, instead of (18), the form

$$-F_{SC}=\left(\frac{f'^2-2xff'}{f^2}\right)'+(x^2)' \tag{45}$$

in this particular situation. The true force is found from the second term of (45). So, by applying the same logic as before, one can insist that the polynomial $f$ should satisfy

$$f'^2-2xff'=c_n f^2, \tag{46}$$

with $c_n$ as an arbitrary constant for the $n$th state, so that the first term in (45) provides a vanishing contribution. But, one can check that no such polynomial is possible. In fact, it is apparent from (46) that while the left side does not have a constant term, the right side does. In other words, a constant $f$ is the *only* possibility. This justifies why nodes do not have a natural place in semiclassical approaches. The role of force in deciphering the problem is noteworthy.

**7. Concluding remarks**

In summary, we intended to touch upon a few areas of approximate calculations for stationary states where force plays some positive role. We hope to have achieved the goal. A few more concrete numerical demonstrations would surely establish $\mu'$ as an effective criterion, either in measuring the goodness of an approximate $\Psi$ or in formulating a variational principle. Finally, we have not considered here the role of force in *perturbation theory*, another standard method of getting approximate stationary states. This will hopefully be considered in some future work.